# PROBING A LABEL-FREE LOCAL BEND IN DNA BY SINGLE MOLECULE TETHERED PARTICLE MOTION

Annaël Brunet[1,2,3,4,+], Sébastien Chevalier[3,4,+], Nicolas Destainville[1,2], Manoel Manghi[1,2], Philippe Rousseau[5,6], Maya Salhi[5,6], Laurence Salomé[3,4], Catherine Tardin[3,4,*].

1 CNRS ; LPT (Laboratoire de Physique Théorique), UMR UPS-CNRS 5152, 118 route de Narbonne, F-31062 Toulouse, France.

2 Université de Toulouse, UPS; LPT, F-31062 Toulouse, France

3 CNRS; IPBS (Institut de Pharmacologie et de Biologie Structurale), UMR UPS-CNRS 5089, 205 route de Narbonne, F-31077 Toulouse, France

4 Université de Toulouse, UPS ; IPBS , F-31077 Toulouse, France

5 CNRS ; LMGM (Laboratoire de Microbiologie et Génétique Moléculaires), UMR CNRS-UPS 5100, 118 route de Narbonne, F-31062 Toulouse; France

6 Université de Toulouse ; UPS; LMGM; F-31062 Toulouse; France

* To whom correspondence should be addressed. Tel: +33 561175468; Fax: +33 561175994; Email: tardin@ipbs.fr
+ The authors wish it to be known that, in their opinion, the first two authors should be regarded as joint First Authors.

## ABSTRACT:
Being capable of characterizing DNA local bending is essential to understand thoroughly many biological processes because they involve a local bending of the double helix axis, either intrinsic to the sequence or induced by the binding of proteins. Developing a method to measure DNA bend angles that does not perturb the conformation of the DNA itself or the DNA-protein complex is a challenging task. Here, we propose a joint theory-experiment high throughput approach to rigorously measure such bend angles using the Tethered Particle Motion (TPM) technique. By carefully modeling the TPM geometry, we propose a simple formula based on a kinked Worm-Like Chain model to extract the bend angle from TPM measurements. Using constructs made of 575 base-pair DNAs with in-phase assemblies of 1 to 7 6A-tracts, we find that the sequence $CA_6CGG$ induces a bend angle of 19° ± 4°. Our method is successfully compared to more theoretically complex or experimentally invasive ones such as cyclization, NMR, FRET or AFM. We further apply our procedure to TPM measurements from the literature and demonstrate that the angles of bends induced by proteins, such as Integration Host Factor (IHF) can be reliably evaluated as well.

## INTRODUCTION

DNA bending was first revealed in the mid-80's on the mitochondrial DNA of trypanosomatid parasites, the kinetoplast DNA (kDNA) (1) and attributed to the intrinsic bending property of the A-tracts sequences present in kDNA (2, 3). These A-tracts were not only abundantly found in other prokaryotic and eukaryotic organisms but they were also shown to have a biological role, for example, by participating in the regulation of transcription (4–13). The binding of protein to DNA, that occurs in most of DNA-related biological processes, was also observed to induce the local bending of DNA (3, 14–16).

As a result, for both intrinsic, sequence-dependent, or protein-induced bending of DNA, a large amount of work has been carried out to characterize it on structural and thermodynamical grounds (15, 16). It is now commonly accepted that the bendability of specific DNA sequences relates to their capacity to be bent under the action of DNA-binding proteins. The bendability of these sequences may stem from their intrinsic bend, their low bending modulus or some specific breathing behavior of the DNA duplex structure that would facilitate its interactions with proteins. A DNA-analysis server based on the bending propensities of tri-nucleotides, that were deduced from DNase I digestion data (17), can be used to predict DNA structure from sequence and get an estimation of the foreseen DNA bend angles (18). However, characterizing rigorously the local bending of DNA molecules is a crucial issue that remains highly challenging (13, 19, 20).

To probe the DNA bendability experimentally, the most popular but complex technique remains the DNA cyclization method which provides a measure of the efficiency of cyclization of DNA fragments in presence of DNA ligase (21, 22). However, this method does not permit to distinguish between changes in bending modulus that could be due to either permanent or transient structural defects and the presence of a local bend, a question for which other experimental strategies are needed. Another indirect approach consists in carrying out gel shift electrophoresis experiments using DNA molecules with several intrinsically bent sequences in-phase and in opposition of phase or circularly permuted DNA fragments (23–25). Though easy to handle, this technique can only provide rough estimates of bending angle and may be difficult to employ in all cases. Both NMR and X-ray require expensive equipment and tedious sample preparation and analysis procedures. They are therefore not routinely employed to detect and quantify the angle of an unknown DNA local bend but to bring details at the base-pair (bp) scale of a DNA structure already known to exhibit a local bend (26, 27). Bend angles have also been deduced from more indirect techniques based on distance-dependent processes such as FRET (28, 29) or Plasmon resonance coupling of nanoparticles (30). Note that for these four last methods, the investigations are restricted to DNA molecules much shorter than the persistence length. Bending angles can also be extracted from single-molecule force-extension curves at large forces, as first proposed in (31). In this approach, the bending angle is inferred from the apparent persistence length, itself being a parameter used to fit experimental force-extension curves. However, extracting accurate values of the persistence length in this context presents several inherent difficulties, notably the fact that the fitting equation is supposed to be valid in the large-force regime where non-linear stretching should also be taken into account (32). More recently direct visualizations by AFM (33, 34) and cryo-electron microscopy (35) have given quantitative measurements of the bend angles. However these methods have limitations. Though AFM apparently gives the most direct access to the bend angles, the measurements are potentially biased by the sample preparation (36, 37). By contrast, the technically demanding cryo-EM is supposed to preserve close-to-native state of the DNA complex, but it may also induce biases due especially to the sample confinement into a 50 nm thick layer (35).

In the present work, we propose a physical method for the measure of DNA bend angles in a single DNA molecule and DNA-protein complex, which leaves intact the DNA conformation by ensuring minimal interaction with surfaces or tagged particles. This method combines our recently developed High Throughput Tethered Particle Motion (HT-TPM) technique and analytical modeling. HT-TPM enables the tracking of the conformational dynamics of hundreds of single DNA molecules in parallel, free to fluctuate in solution (38) (Fig. 1A, SI video). To extract the bend angles from HT-TPM data, we developed a simple analytical formula based on a kinked Worm-Like Chain (WLC) model that we validated on simulated data. Applied to constructs made of 575 bp DNAs with in-phase assemblies of 1 to 7 6A-tracts, we find that the sequence $CA_6CGG$ has an intrinsic bend angle of 19° ± 4°. In addition, the slight difference between our experimental data and our analytical model for a regular DNA suggests that even a DNA molecule with a randomly chosen sequence may contain a global curvature. We further apply our procedure to TPM measurements from the literature and demonstrate that the angles of bends induced by proteins, such as Integration Host Factor (IHF) can be reliably evaluated as well.

## MATERIALS AND METHODS
### DNA constructs
DNA molecules were produced by PCR (oligonucleotides (Sigma-Aldrich): Biot-F575 ATAAGGGCGACACGGAAATG and Dig-R575 CGTGCAAGATTCCGAATACC) on pTOC plasmids, derived from pBR322 (39). Synthetic (GeneScript) DNA molecules (88 bp) containing increasing number of A6-tracts (0 to 7) were synthesized (Fig.1B) and were inserted between the HindIII and SalI restriction sites of pBR322. PCR products were purified as described in Diagne et al. (40).

**HT-TPM setup and procedure**
HT-TPM on chip assembly and the experimental setup are such as described in (38). The DNA/particle complexes were visualized in TPM buffer with an ionic strength of 165 mM (1.06 mM $KH_2PO_4$, 3.00 mM $Na_2HPO_4$, 154 mM NaCl, 1 mg.mL$^{-1}$ Pluronic F127, 0.1mg.mL$^{-1}$ BSA) at 21 ± 1°C using a dark-field microscope (Axiovert 200, Zeiss). Acquisitions of 5 min were performed at a recording rate of 25 Hz with acquisition time of 40ms on a CMOS camera Dalsa Falcon 1.4M100. The field of observation covers an area of ~ 215 µm x 160 µm. Details can be found in SI text.

The software Nanomultiplex co-developed with Magellium Toulouse (request should be addressed at info@magellium.fr) tracks in real time the positions of all the particles using the centroid method, averages these absolute positions on a 5 s window giving access to the anchoring point of the DNA molecule, calculates the 2D-vector positions of the bead $\vec{R}_{exp\|raw}$ relative to the anchoring point of the DNA which corrects for experimental drift, calculates the asymmetry factor of the particle trajectories (41), and the amplitude of motion of the particle defined as $\sqrt{\langle \vec{R}^2_{exp\|raw}\rangle}$ and noted $R_{exp//raw}$. In a general manner, we will use indifferently $\sqrt{\langle \vec{R}^2 \rangle}$ and $R$ in the following. The averages, performed in the calculation of the asymmetry factors and amplitudes of motion, are taken over a sliding window of 5 s along the time trace. We invite the reader to refer to (42) for the detailed calculations of $R_{exp//raw}$.

Finite exposure time of detectors, $T_{ex}$, equal here to 40 ms, can lead to a blurring effect in single molecule (or particle) tracking experiments, as investigated for example in (42). The correlation time of the positions $\tau_\|$, of about 20 ms, was calculated for each DNA-particle complex and then injected in the following equation

$$R_{exp\|} = R_{exp\|raw}\left[2\frac{\tau_\|}{T_{ex}} - 2\left(\frac{\tau_\|}{T_{ex}}\right)^2\left(1 - e^{-\frac{T_{ex}}{\tau_\|}}\right)\right]^{-1/2} \quad (1)$$

to correct the amplitudes of motion recorded on each trace from blurring. In order to quantify the small differences expected on $R_{exp//}$, we set up a two-step procedure that is described in detail in the SI text.

**TPM simulations**
We performed Kinetic Monte Carlo simulations on the particle-DNA complex to predict the particle to anchor 2D-distance. We invite the reader to refer to (42) for the details of these Kinetic Monte Carlo simulations.

The bent sequences used in the experiments are simulated by setting a fixed angle between three successive monomers located at the center of the DNA molecule. A full range of angles were studied in successive simulations: 0, 18, 30, 45, 50, 60, 72, 90, 120 and 180 degrees. The 2D-vector of the particle position $\vec{R}_{sim\|}$ is measured throughout simulations and utilized to estimate the amplitude of motion defined as $\sqrt{\langle \vec{R}^2_{sim\|}\rangle}$, the average being taken along the trajectory.

**Influence of the particle size on the angle measurements.**
The particle size cannot be known exactly. Indeed, on the one hand, there is a limited control of their radius by the manufacturer leading to an uncertainty of 3 nm. On the other hand, the subsequent functionalization of the particles confer them a slightly larger radius $R_p$ (Fig. 1A). Indeed, the layer of antibody used to connect the DNA molecule to the particle is expected to be a few nanometers wide. Since in Eq. 2, $R_p$ precisely refers to the distance between the particle center and the extremity of the DNA molecule, and being not able to exactly infer it, we chose to explore two cases: $R_p$ = 150 nm, previously used, and $R_p$ = 155 nm, the real values probably dwelling between both values.

For $R_p$ = 155 nm, $\theta_1$ was found to be equal to 17° ± 2°, while for $R_p$ = 150 nm, $\theta_1$= 15°± 2° . The bend angle values obtained considering or not a 5 nm increase of the effective radius of the particles in this precise case cannot be distinguished. Though particles with a well-defined size would lead to a more precise bend angle, the uncertainty on the exact value of $R_p$ does not appear to be a critical issue in a typical case such as the one described here.

## RESULTS
### A new experimental strategy

The presence of a localized bent structure within a DNA molecule is expected to induce the reduction of the apparent end-to-end distance of the entire DNA molecule. Such an effect will be much easier to detect on short DNA molecules. Such an effect will be much easier to detect on short DNA molecules. To measure it, we chose 575 bp long DNA molecules that were immobilized on a functionalized coverslip by one end, and attached to a 150 nm radius particle at their other end permitting their video tracking (Fig. 1A). This experimental design represents a good compromise for an easy particle detection and a nearly force-free measurement, as the effective force exerted by the particle on the DNA is only a fraction of pN (43). In that way, the 2D projection of the particle displacement relative to the anchoring point of the DNA molecule gives access to its root-mean-squared end-to-end distance projected on the grafting surface, noted $R_{exp//raw}$, which depends on the length and the conformational state of the monitored molecule. $R_{exp//raw}$ was corrected for the blurring effect, caused by the acquisition system, to obtain $R_{exp//}$, using Eq. 1 of *Material and Methods*. TPM is capable of revealing changes in tether length equivalent to an apparent contour length as small as 100 bp (42, 44). To obtain a good precision, inferior to 1 nm, a large amount of experimental data (see table 1) is required because of the intrinsic dispersion of data due to an unavoidable variability of the DNA-particle and DNA-substrate links. To do so, we take advantage of our recently developed biochip that permits us to accumulate acquisitions on several hundreds of single DNA molecules in parallel by HT-TPM with a typical error on $R_{exp//}$ equal to 0.4 nm (38).

In order to evaluate the capacity of this technique to detect and quantify local bending angles, we produced a series of DNA molecules based on a unique plasmid series. It incorporates a central 88 bp region, smaller than the DNA persistence length ($L_p \approx 150$ bp), containing 1 to 7 CA$_6$CGG sequences in-phase or in opposition of phase (Fig. 1B and *Material and Methods*). We chose the CA$_6$CGG sequence, known to be a sequence inducing a large bend (45). The in-phase A-tracts are located every integer number of helix turns, whereas those in opposition of phase are located every half-integer number of helix turns (Fig 1B).

By HT-TPM, we measured $R_{exp//raw}$ and calculated $R_{exp//}$, the amplitude of motion, for this series of 575 bp long DNA molecule (Table1).

### A kinked Worm-Like Chain model to analyze TPM data

Obtaining the bend angle from $R_{exp\parallel}$ requires the calculation of the root-mean-square end-to-end distance of the DNA molecule, $R_{DNA}$, and an appropriate theoretical model giving the variation of $R_{DNA}$ as a function of a local bend angle, denoted by $\theta$.

To validate this analytical tool, we simulated a DNA-particle complex in a TPM setup (see *Material and Methods and SI text)* where a bend of fixed angle $\theta$ varying from 0 to 180° was incorporated in the middle of the DNA molecule. The corresponding projection of the mean-square end-to-end distance $R_{sim\parallel}$ was computed accordingly, as sketched in Fig. 1A.

$R_{DNA}$ was extracted from these numerical results by correcting for the effects of the particle and of the glass substrate (42, 43). To achieve it, we explored two strategies. The first one is a minimal model in which the particle and the DNA molecule are considered statistically independent and the effect of the substrate is ignored. This last assumption is correct when the particle is very small, $R_P \ll R_{DNA}$, and the DNA molecule is either very long ($L \gg L_p$) or very short ($L \ll L_p$), leading to $\langle R_\parallel^2 \rangle = \frac{2}{3}\left[R_{DNA}^2 + R_p^2\right]$ and thus to:

$$R_{DNA} = \left[\tfrac{3}{2}\langle R_\parallel^2 \rangle - R_p^2\right]^{1/2} \qquad (2)$$

However, it has been shown that the effect of the particle can matter, because the hard-core interaction between the particle and the substrate reduces the number of degrees of freedom accessible to the molecule. The second strategy relies on a more sophisticated protocol proposed by Segall et al. in (43) to correct for the effects of the particle and of the glass substrate for very long DNA. We refer to this approach as Segall's method. It consists in solving the equation

$$\tfrac{3}{2}\frac{\langle R_\parallel^2 \rangle}{\langle R_{DNA}^2 \rangle} = 1 + \frac{2N_R}{\sqrt{\pi}\,erf(N_R)} \qquad (3)$$

where *erf* is the error function and $N_R = \sqrt{6}R_p/\sqrt{\langle R_{DNA}^2 \rangle}$. To use this Eq. 3 in the present context, we crudely extrapolated the results of (43), obtained for the Gaussian chain case ($L \gg L_p$), to the semi-flexible regime ($L \sim L_p$), by replacing $2LL_p$ by $\langle R_{DNA}^2 \rangle$. Note that Eq. 2 is nothing but the expansion at order 1 in $N_R$ of Eq. 3.

To obtain the bending angle, we fit the corrected data using a WLC model (46) on a polymer of length $L$ with persistence length $L_p$, with a bend located at distance $l$ from one end which locally induces a curvature with an angle $\theta$ (see Fig. 1A). The mean-squared end-to-end distance is given by (SI text):

$$\langle \vec{R}^2 \rangle(\theta) = 2L_P^2\left[\frac{L}{L_P} - 2 + e^{-\frac{l}{L_P}} + e^{-\frac{L-l}{L_P}} + \cos(\theta)\left(1 - e^{-\frac{L-l}{L_P}} - e^{-\frac{l}{L_P}} + e^{-\frac{L}{L_P}}\right)\right] \qquad (4)$$

For *L*=575 bp and *l*=L/2, one derives from Eq. 4 the fitting formula, $R_{\mathrm{WLC}}(\theta) = \sqrt{\langle \vec{R}^2 \rangle} = D\sqrt{1 + 0.342\cos\theta}$. In Fig. 2, both methods were applied to the numerical data. For example, for $\theta = 0$, the true WLC value of $R_{DNA}$ is $R_{WLC}(\theta = 0)$= 121.9 nm (see *Material and Methods*), in-between the values obtained with Segall's model and the minimal one. Whereas the minimal method underestimates $R_{WLC}$, Segall's one overestimates it, both methods leading to comparable relative errors of about 10%. We thus choose $D$ as a free fitting parameter in order to account for an offset at $\theta = 0$. One observes in Fig. 2 that whereas the minimal method gives a very satisfying fit, Segall's one leads to a worse one. Hence extrapolating Segall's calculation to the case of semi-flexible DNA molecules including a local bend appears to be less adapted than the minimal model. We thus use the minimal model in the remainder of this paper.

**Bending angle measurements on a regular DNA**

For the set of DNA molecules containing series of in-phase 6A-tracts (Table 1), we found by HT-TPM that $R_{exp//}$ diminishes from 149.7 nm to 140.4 nm. When the initial 88 bp internal fragment of the DNA molecules (6An0) was replaced by a fragment containing only one (6An1$^P$), two (6An2$^P$), or three 6A-tracts (6An3$^P$) in-phase, $R_{exp//}$ remained in a range of 0.6 nm from its initial value (no 6A-tract), which appears to be very close to the 0.4 nm incertitude range we estimated for our HT-TPM measurements according to the method described in the *Material and Methods* section. $R_{exp//}$ decreased down to 146.5 nm for DNA molecules with four 6A-tracts in-phase (6An4$^P$), 145.2 nm with six 6A-tracts in-phase (6An6$^P$), and 140.4 nm with seven 6A-tracts in-phase (6An7$^P$). These decreases in $R_{exp//}$ might stem from an intrinsic bend of the 6A-tract or a decrease in bending modulus of this sequence. In this latter case, one would expect the insertion of four 6A-tracts (6An4$^O$) in opposition of phase to lead to a decrease in $R_{exp//}$ similar to the one measured on 6An4$^P$. In fact, 6An4$^O$ showed no decrease in $R_{exp//}$ as it was found equal to 151.2 nm. Taken together, these results prove that, beyond four 6A-tracts, $R_{exp//}$ decreased when the number of 6A-tracts in-phase increases due to an intrinsic bend of the 6A-tract.

In order to calculate $R_{DNA}$, we injected these $R_{exp//}$ values in Eq. 2. We assume then that each of the *n* successive 6A-tracts inserts in-phase imposes the same bending angle $\theta_1$ and postulate as a first order simplification hypothesis that $\theta = n\theta_1$. With $L = 575$ bp, $l = 318$ bp, we now obtain the following equation:

$$R_{WLC}(n) = D[1 + 0.338 \cos(n\theta_1)]^{1/2} \qquad (5)$$

The fits of $R_{DNA}$ shown in Fig. 3 (black symbols and fits for $R_p = 150$ nm) are reasonably good leading to $D = 92 \pm 2$ nm and $\theta_1 = 15° \pm 2°$ for the value of the bend angle of each CA$_6$CGG sequence. In the *Material and Methods* section, we studied the influence of small variations of the particle size on this result and found comparable results within error bar.

**Influence of a global DNA curvature on the angle measurements**

The experimental values of Fig. 3 suggested that $R_{DNA}$ might be non-monotonous when *n* increases, contrary to what is expected from Eq. 5. One reason for this observation could be the existence of an intrinsic curvature of the DNA molecule. To account for it, we have chosen to refine our analysis and use the more general fitting form:

$$R_{WLC}(n) = D[1 + 0.338 \cos(n\theta_1 - \theta_0)]^{1/2} \qquad (6)$$

where $\theta_0$ mimics the whole intrinsic bend of the $n = 0$ molecule. For sake of simplicity, we implicitly assume that it is accumulated at the same position as the bends when $n > 0$ (17, 23). The resulting fits, displayed in Fig. 3 (dotted lines), are slightly better than before.

To sum up, the proposed refinement suggests that the true bending angle $\theta_1$ is 19° ± 4° per bend, while the global bend of the DNA molecule would have an angle $\theta_0$ between 25° ± 20°. Although the fit including a global curvature of the DNA molecule seems better when looking at Fig. 3, this new parameter $\theta_0$ can only be determined with a poor precision. This is due to the fact that this angle is actually delocalized on the whole 575 bp sequence. A more sophisticated model is needed to take this point better into account.

**Influence of the bending modulus of the bent sequence on the angle measurements.**

Alternatively, it was in principle possible that the increase in the DNA end-to-end distance observed at small values of *n* was associated with an increase of the molecular rigidity of the inserts. We analytically derived the corrections to $R_{DNA}$ due to an insert of finite length and of increased rigidity using the WLC model (SI text). At small insert length, the increase is linear with the insert length (Fig. S1). At first order, this effect is additive with the effect of a bend. The corresponding fits are displayed in Fig. S2 and Table S3. The obtained values of $\theta_1$ were close to the previous ones, thus showing that this alternative approach does not differ significantly from the previous one. This could be anticipated since a linear expansion of Eq. 6 adds a linear term into Eq. 5 if one assumes that $\theta_0 \ll 1$. As a result, the combined HT-TPM/kinked-WLC approach might not easily discriminate between an increase of rigidity of the inserts and an intrinsic bend. However, when fitting the data of the DNA molecules, containing 0 to 4 6A-tracts inserts, with this additional linear term, we found a linear increase of the amplitude of motion of 1.9 nm per 10bp 6A-tract. It would correspond to a persistence length of about 89 nm for this bent sequence. There exists another way to evaluate the rigidity contribution of the 6A-tracts. It consists in considering the data obtained for 6An0 (no 6A-tracts) and 6An4[O] (four 6A-tracts in opposition of phase). An increase in $R_{DNA}$ of 3.2 nm (105.4 nm versus 108.6 nm; Table 1) was then measured and gave access to a persistence length equal to 129 nm for this 55 bp sequence by using the formula in SI text. These persistence lengths are estimated with a low precision due to the poor sensitivity of $R_{DNA}$ to an increase of $L_p$ above 150bp (Fig. S4). In any case, whatever the way used to estimate the persistence lengths of these 6A-tract sequences, values were very high compared to the 50 nm measured for random double strand DNA in the present salt conditions (47). As a result, an increase in rigidity of the 6A-tracts could not account for our data, and introducing a global bend $\theta_0$ is probably more adapted to this case.

## DISCUSSION
**DNA bending can be measured using HT-TPM**
In this paper, we develop a procedure to measure a bend angle localized inside long DNA molecules with a good accuracy. For that, we built a platform plasmid based on pBR322 that enabled us to generate several 575 bp long DNA molecules with various 88 bp long DNA constructs close to their center. We used 6A-tracts as a model of bent sequences and showed that, by assembling them every 10.5 bp, we could amplify the observed decrease in $R_{exp//}$ as a function of the number of bent sequences. Furthermore, making use of the HT-TPM, we are able to reliably detect variations of $R_{exp//}$ down to 2%. Using Eq. 2 and Eq. 1, we corrected experimental raw data for the effects of both camera averaging and particle radius and obtain $R_{DNA}$, which was correctly fitted using a kinked WLC model (Eq. 4). We thus got access to the bend angle of the chosen sequence $\theta_1 = 19° \pm 4$.

To sum up, this entire procedure can be readily applied to quantify the bend related to a specific DNA structure together with an evaluation of the intrinsic curvature of the entire DNA molecule. To do so, our platform plasmid can be used to build an *ad hoc* series of DNA molecules with increasing numbers of the DNA sequences under study assembled in-phase. HT-TPM measurements have first to be corrected for the camera blurring effect using Eq. 1 before being injected in Eq. 2 to get the DNA end-to-end distance. The resulting series of data, DNA end-to-end distance as a function of the number of assembled sequences, are adjusted with Eq. 4 where an intrinsic bend angle is incorporated if necessary.

**Comparison with other methods for the measurements of CA$_6$CGG bend angles**

Crothers *et al.* combined numerical simulations and cyclization experiments on DNA fragments of 105 and 210 bp lengths to get the relative bend induced by this sequence (48). They found a bend angle equal to 19° ± 2°. It should be stressed that cyclization is an indirect method which (i) enforces the looping of short DNAs ($\approx L_p$) that can also be due to kinks or small denaturation bubbles (22), (ii) requires numerical simulations of the specific sequence with a large number of unknown parameters for the fitting procedure. In our case, we need at most three free parameters, namely $D$, $\theta_0$ and $\theta_1$.

More precise results were obtained by MacDonald *et al.* who used NMR spectroscopy on partially aligned DNA molecules (26). In particular, they found that the overall helix axis of the DNA dodecamer GGCA$_6$CGG exhibits a bend of 19°± 1° towards the minor groove of the A-tract. In addition, NMR results confirmed the major role played by the joint located between the 6A-tract and the next CG sequence that was found to be responsible for 14° out of the 19° of the global bend angle of the dodecamer. NMR is limited to small DNA fragments and cannot be easily extended to DNA-protein complexes.

X-ray crystallography showed A-tracts structures with various conformations due to the crystal constraints (49). As it is complicated to determine the most representative one, we did not try to compare X-ray crystallography results to ours.

Experiments based on FRET also gave access to the bend angle value of the CA$_6$CGG-CGA$_6$CGG-CA$_6$CGG sequence prepared in solutions with a broad range of salt concentration (28). The bend angle was deduced from the average dye-to-dye distance between two fluorophores located at the ends of the 31 bp DNA fragment. It was shown to vary from 23° ± 4° to 41° ± 4° with NaCl concentration increasing from 10 to 500 mM. At the ionic strength of 165 mM we carried out the experiments, Tóth *et al.* found a bend angle ranging between 36° ± 4° and 41° ± 4° which is in slight disagreement with the 3 x (19° ± 4°)= 57° ± 7° that we obtain. This discrepancy could be due to the sequence located on the middle which is slightly different with the flanking C being replaced by G. According to Koo *et al.* (45), the middle sequence should induce a bend smaller, up to 10 %, than the two other ones. As a result, the total bend angle of the Toth's sequences is necessarily smaller than the one obtained by a 3-fold assembly of our sequence. Another source of discrepancy is that distance measurements by FRET are highly sensitive to the precise geometry of the attachment of the dyes at the extremity of the DNA molecules which is difficult to determine accurately. Moreover, the dyes are very close to the studied sequence and may interfere with it.

Finally, direct imaging of 2 to 8 6A-tracts assembled in-phase has been made by AFM (33). By measuring the planar end-to-end distance of these molecules, it was deduced a mean bend angle for each 6A-tract of about 13.5° with a precision of about 1°. The high precision that was obtained here does not take into account the possible bias induced by the mechanical constraints exerted onto the DNA molecule deposited on mica and visualized in air. Indeed the divalent ions that stick the DNA to the surface are likely to modify significantly the electrostatic interactions and therefore the DNA elastic properties (37).

Using HT-TPM to characterize the 6A-tracts enabled us to obtain values in reasonable agreement with those measured in these preceding studies by other methods. Besides, our measurements were carried out on DNA molecules containing the bent inserts far from the surface and the labeling particle, which thus do not interfere with the insert under examination. This is not possible with the previously mentioned techniques. Though very informative, most of them remain limited to the study of short DNA fragments. Furthermore, our underlying model is simple, appealing to elementary and robust polymer physics arguments, whereas interpreting data incoming from NMR, X-ray scattering, FRET or cyclization experiments is far more indirect and tedious. Finally, parallelizing single molecule experiments is a pre-requisite here, the very small variations of the amplitude of motion that we intend to monitor requiring intensive sampling.

**DNA heterogeneity and mechanical characterization**
As expected the end-to-end distance of 6An0, calculated as $R_{\text{DNA}}(\theta = 0)$, is found notably smaller, 107.7 ± 7.1 nm, than the expected WLC value, namely 121.9 nm. It suggests that the WLC model is not perfectly adequate to the DNA sample studied here. More importantly, whatever the equation we used to fit our HT-TPM data, the fits do not match exactly the experimental data within error bars. This can be due to an under-estimation of our confidence intervals. Even though increasing our sample sizes, we did not manage to reduce this discrepancy. We propose that it could be due to the intrinsic bend accumulated along the molecule, the additional effect of which would reduce the overall apparent end-to-end distance. Notably, when deleting some random DNA regions and replacing them by bent inserts, we conjecture that the deleted regions already bore some small but non-zero curvature, the effect of which can be either compensated or accentuated when replaced by an insert. This leads to small corrections of the end-to-end distance around the average trend described by Eq. 4. This work thus highlights that a given sequence generically involves a small but non-vanishing intrinsic bend to DNA molecules, the amplitude of which typically grows like the square root of the molecule length, owing to the central-limit theorem, because the small bending angles related to individual base pairs add up independently along the chain. From a statistical physics perspective, our results point to the need to adapt statistical models to incorporate the fact that typical DNA molecules with a "random" sequence adopt a curved shape rather than a straight one in the rigid regime $L<L_p$ (13).

**Quantification of protein-induced DNA bends by HT-TPM**
TPM is one of the very few methods permitting the monitoring of DNA looping. In addition, relying on the theoretical framework proposed here, we claim that HT-TPM can also reveal the protein-induced bending of the DNA interaction site and give access to the protein-induced bend angle.

When studying the interaction of a protein with a DNA molecule, a decrease in $R_{DNA}$ may be observed after the addition of the protein on DNA molecules containing only one site of interaction (40, 44, 50–52). In this case, we can still evaluate the bend angle $\theta$ from the two DNA end-to-end distance values obtained before and after addition of the protein by using Eq. 7, derived from Eq. 4 assuming that the global intrinsic bend $\theta_0$ is negligible,

$$cos(\theta) = 1 + \frac{e^{-\frac{L}{L_P}}+\frac{L}{L_P}-1}{1-e^{-\frac{L-l}{L_P}}-e^{-\frac{l}{L_P}}+e^{-\frac{L}{L_P}}}\left(\frac{R_{DNA}^2(\theta)-R_{DNA}^2(0)}{R_{DNA}^2(0)}\right) \qquad (7)$$

Note that if an intrinsic angle $\theta_0$ was present before insertion of the sequence of interest, this formula gives access to the angle variation $\theta - \theta_0$ instead of $\theta$.

As a test of its accuracy, we applied this simplified method to 6An4[P], 6An6[P] and 6An7[P] and obtained a mean value of one 6A-tract insert equal to 15° ± 2°, the uncertainty is here calculated as the standard deviation of the data obtained for the three DNA molecules (Table S5). This value is in good agreement with the value 15° ± 2° obtained from the fit performed on the entire set of data with $\theta_1$ as the only free angle parameter ($\theta_0 = 0$). In addition, using Eq. 7 and considering equal to 2% the minimal variation in $R_{exp//}$ that can be detected with HT-TPM on this 575 bp DNA molecule, one can only measure bend angles larger than 33°. To measure smaller bend angles, the construction of a series of repeated sequence is required.

On the ground of this simple formula (Eq. 7), we analyzed the TPM results obtained by Mumm et *al.* (52) concerning the bend induced by the Integration Host Factor (IHF) that is known to bind to some specific binding sites and play thus a major role as architectural protein in prokaryotes. Upon binding of the protein to a single site of a *L*=1943 bp long DNA, the molecule behaved as a 1659 bp long DNA which corresponds to $R_{DNA}$ decreasing from $R_{DNA}$ =249 nm to 229 nm. The binding site being located at *l*=301 bp away from one extremity, we evaluate the IHF induced-angle to be about 180° using Eq. 7. This value is in very good agreement with the crystallography measurements that show that IHF provokes a bend with an angle of at least 160° (53), especially as it should be considered that the end-to-end distance tends to saturate for such high angles.

Our analytical models consider separately the cases of a bend with a fixed angle and of a local flexible hinge, while a great number of DNA-binding proteins are now considered to induce both effects. To account for these two mechanical changes occurring simultaneously as well as the extended deformation of the protein-DNA interaction site, a more precise theory is needed. Nevertheless, we believe that our method could still be applied to those cases and give access to some valuable though less precise information.

We can therefore conclude that the combined HT-TPM/kinked-WLC approach provides an efficient method to estimate the angle of fixed local DNA bend either intrinsic to a sequence or induced by the binding of proteins. With many advantages over the existing methods, we believe that our approach will permit the refined characterization of DNA geometry in various contexts and to shed new light on DNA-protein complexes.

## SUPPORTING INFORMATION

Supporting information is available at NAR Online.

## ACKNOWLEDGEMENTS

We thank François CORNET for his helpful discussions on the DNA construction.
We belong to the GDR CellTiss.

## FUNDING

This work was supported by CNRS, University of Toulouse 3 and ANR-11-NANO-010 'TPM On a Chip'.

*Conflict of interest statement.* None declared.

## TABLE AND FIGURE

Table 1. Amplitude of motion, $R_{exp\parallel}$, corrected from the blurring effect and end-to-end distance, $R_{DNA}$, extracted with a particle radius $R_p = 150$ nm for the set of constructs.

| DNA samples | Number of trajectories | $R_{exp\parallel}$ (nm) | Uncertainty (nm) | $R_{DNA}$ (nm) |
|---|---|---|---|---|
| 6An0 | 3496 | 149.7 | 0.2 | 105.4 |
| 6An1$^P$ | 2728 | 149.1 | 0.2 | 104.1 |
| 6An2$^P$ | 2354 | 149.7 | 0.2 | 105.4 |
| 6An3$^P$ | 2904 | 149.6 | 0.2 | 105.2 |
| 6An4$^P$ | 2604 | 146.5 | 0.2 | 98.5 |
| 6An6$^P$ | 348 | 145.2 | 0.4 | 95.5 |
| 6An7$^P$ | 431 | 140.4 | 0.4 | 84.1 |
| 6An4$^O$ | 2990 | 151.2 | 0.2 | 108.6 |

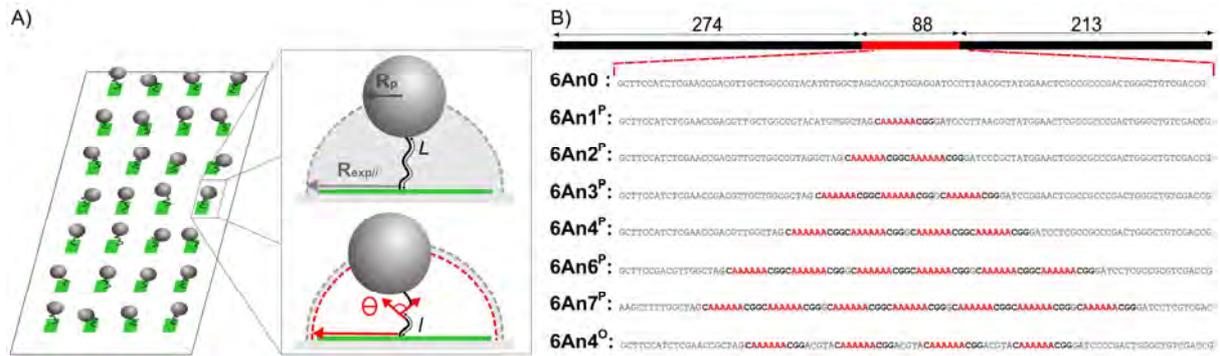

Figure 1. A) Schematic HT-TPM setup to measure the apparent length of a DNA molecule. The amplitude of motion, $R_{exp||}$, of a particle tethered to a surface by a DNA with a contour length $L$ depends on its effective length which varies with the angle $\theta$ of a bend located at distance $l$ from one end chosen nearby the center of the molecule. For high-throughput measurements, individual DNA-particle complexes are immobilized on an array of functionalized sites. See *Material and Methods* for details.

B) Representation of a typical 575 bp long DNA fragment used in this study with the variable 88 bp central region (Sizes are indicated in bp, details in SI). The central region contains $n$ copies of a $CA_6CGG$ sequences (in red), which are predicted to be bent (see text). The sequences of the 88 bp long insert of the 8 DNA molecules studied here are represented with the $CA_6CGG$ sequences. Each sequence is named as a function of the number of A contained in the $CA_6CGG$ (6A), followed by the number of repeats (nX) and the phasing (in exponent, P: in-phase, O: in opposition of phase).

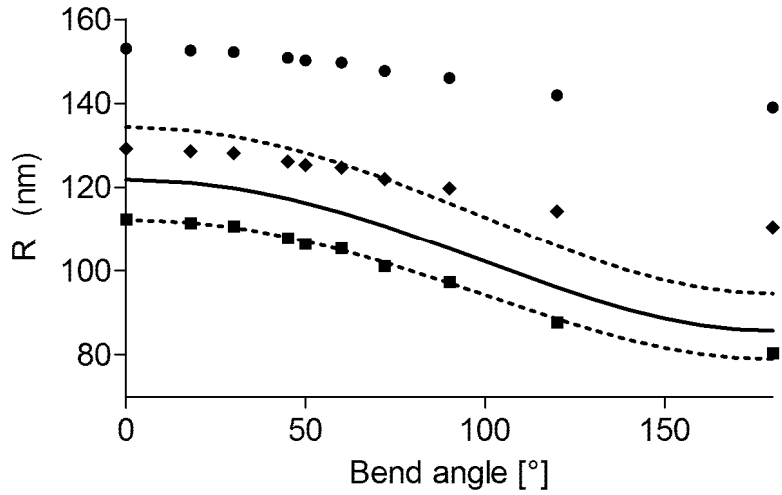

Figure 2. Simulated HT-TPM amplitudes of motion $R_{sim//}$ (●), together with the corrected values $R_{DNA}$ obtained through Segall's method (◆) and the minimal one (■); see text for details. The DNA molecule length is $L$ = 575 bp and the particle radius is $R_p$ = 150 nm, as in the experiments. Fits of the numerical data, corrected by both the Segall's method and the minimal one, are shown as dotted lines using Eq. 4. The curve corresponding to the WLC model, with no adjusted parameters, is shown as a solid line.

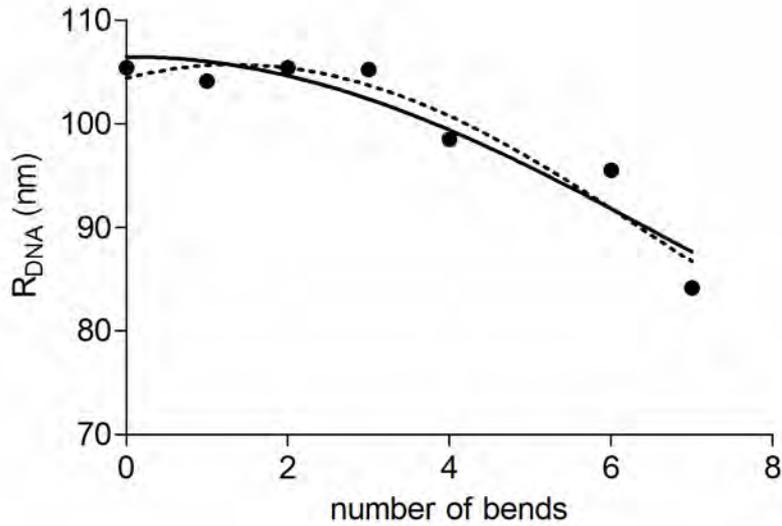

Figure 3. Experimental data after correction of the particle effects for the labelling particle $R_p$ = 150 nm (●) using the minimal model, and two series of fits with D and $\theta_1$ (straight line), or D, $\theta_1$ and $\theta_0$ (dotted line) as free parameters. Error bars are smaller than the symbol size. Using Eq. 5, ones finds: $D$= 92.0 nm, $\theta_1$ = 15° for $R_p$ = 150 nm. Using Eq. 6, one finds $D$ = 91 nm, $\theta_1$ = 19°, $\theta_0$ = 25° (see S3 table for details).

# SUPPORTING INFORMATION

## SI text

### List of abbreviations

$\vec{R}_{exp\parallel raw}$:      2D-vector positions of the particle measured experimentally

$R_{exp\parallel raw}$      Amplitude of motion of the particle defined as $\sqrt{\langle \vec{R}^2_{exp\parallel raw}\rangle}$

$R_{exp\parallel}$      Amplitude of motion of the particle corrected from the blurring effect

$\vec{R}_{sim\parallel}$      2D-vector positions of the simulated particle

$R_{sim\parallel}$      Amplitude of motion of the simulated particle defined as $\sqrt{\langle \vec{R}^2_{sim\parallel}\rangle}$

$R_{DNA}$      End-to-end distance of the DNA molecule

$R_P$      Radius of the labelling particle

$L$      Contour length of the DNA molecule

$\ell$      Distance separating the kink to one extremity of the DNA molecule

$\theta$      Bend angle

$L_P$      Persistence length of the DNA molecule

### DNA constructs

The sequences of the 88bp long insert (Fig. 1) are the following ones:

**6An0**: `GCTTCCATCTCGAACCGACGTTGCTGGCCGTACATGTGGCTAGCACCATGGAGGATCCCTTAACGCTATGGAACTCGCCGCCCGACTGGGCTGTCGACCG`

**6An1$^P$**: `GCTTCCATCTCGAACCGACGTTGCTGGCCGTACATGTGGCTAGCAAAAAACGGGATCCCTTAACGCTATGGAACTCGCCGCCCGACTGGGCTGTCGACCG`

**6An2$^P$**: `GCTTCCATCTCGAACCGACGTTGCTGGCCGTAGGCTAGCAAAAAACGGCAAAAAACGGGATCCCGCTATGGAACTCGCCGCCCGACTGGGCTGTCGACCG`

**6An3$^P$**: `GCTTCCATCTCGAACCGACGTTGCTGGCGCTAGCAAAAAACGGCAAAAAACGGGCAAAAAACGGGATCCGGAACTCGCCGCCCGACTGGGCTGTCGACCG`

**6An4$^P$**: `GCTTCCATCTCGAACCGACGTTGGCTAGCAAAAAACGGCAAAAAACGGGCAAAAAACGGCAAAAAACGGGATCCTCGCCGCCCGACTGGGCTGTCGACCG`

**6An6$^P$**: `GCTTCCGACGTTGGCTAGCAAAAAACGGCAAAAAACGGGCAAAAAACGGCAAAAAACGGGCAAAAAACGGCAAAAAACGGGATCCTCGCCGCGTCGACCG`

**6An7$^P$**: `AAGCTTTTGGCTAGCAAAAAACGGCAAAAAACGGGCAAAAAACGGCAAAAAACGGGCAAAAAACGGCAAAAAACGGGCAAAAAACGGGATCCTCGTCGAC`

**6An4$^O$**: `GCTTCCATCTCGAACCGCTAGCAAAAAACGGACGTACAAAAAACGGACGTACAAAAAACGGACGTACAAAAAACGGGATCCCCGACTGGGCTGTCGACCG`

**Assembly of the HT-TPM biochip**

*Patterning of anchoring sites on functionalized coverslips*

Regular arrays of rhodamine-labelled neutravidin (Molecular Probes) were obtained using a standard micro-contact printing protocol. Briefly, a stamp made of PDMS with squared pillars of 0.8 µm size and 3 µm pitch was inked with a 20 µg/mL neutravidin solution in PBS (Euromedex) for 1 min, then washed with deionized water and dried under nitrogen flow. The stamp was then brought into close contact with epoxidized glass support for 1 min during which the protein is transferred onto the surface.

*Formation of DNA-bead complexes*

Polystyrene carboxylated beads (Merck) of (150 ± 3) nm radius were covalently coated with anti-digoxigenin antibodies (Roche) using EDAC (Sigma-Aldrich), activation and storage buffers (Ademtech). A 100 pM solution of suspended functionalized beads was mixed with an equal volume of a solution of 50 pM DNA bearing a digoxigenin on one end for 20 min at room temperature in PBS buffer supplemented with 1mg/mL of pluronic 127 (Sigma-Aldrich) and 0.1 mg/mL BSA (Sigma-Aldrich), noted TPM buffer. This lead to pre-assembled DNA-bead complexes.

*Assembly of the fluidic observation chamber for HT-TPM experiments*

A 250 µm thick silicone tape was cut and used as a spacer between the patterned coverslip and an epoxidized glass slide with 2 holes for inlet and outlet to obtain a working flow cell. The so-formed analysis chamber was rinsed and incubated with TPM buffer for 30 min at room temperature. The DNA-bead complexes solution was introduced in the flow cell and incubated over night at 4°C.

Prior to visualization, the flow cell was extensively rinsed by injecting 30 chamber volumes of TPM buffer. Then, for each condition, movies of 5 min were recorded on different zones in the same flow cell and analyzed. To ensure reproducibility, experiments were repeated on different days.

**HT-TPM experimental setup**

*Instrumentation for microscopy imaging*

The tethered beads were visualized using a dark-field microscope (Axiovert 200, Zeiss) equipped with a x32 objective and an additional x1.6 magnification lens and acquired for 5 min at room temperature, at a recording rate of 25 Hz and with a duration of acquisition of 40ms, on a CMOS camera Dalsa Falcon 1.4M100. The field of observation covers an area of ~ 215 µm x 160 µm.

*Single particle tracking*

The software Nanomultiplex co-developed with Magellium Toulouse (request should be addressed at info@magellium.fr) tracks in real time the positions of all the particles using the centroid method, averages these absolute positions on a 5 s window giving access to the anchoring point of the DNA molecule, calculates the 2D-vector positions of the bead $\vec{R}_{exp\|raw}$ relative to the anchoring point of the DNA which corrects for experimental drift, calculates the asymmetry factor of the bead trajectories (S) (Blumberg S, Gajraj A, Pennington MW, Meiners J-C (2005) Three-Dimensional Characterization of Tethered Microspheres by Total Internal Reflection Fluorescence Microscopy. Biophys J 89(2):1272–1281), and the amplitude of motion of the particle defined as $\sqrt{\langle \vec{R}^2_{exp\|raw}\rangle}$ and noted $R_{exp//raw}$. In a general manner, we will use indifferently $\sqrt{\langle \vec{R}^2 \rangle}$ and $R$ in the following.

The averages, performed in the calculation of the asymmetry factors and amplitudes of motion, are taken over a sliding window of 5 s along the time trace. We invite the reader to refer to (Plenat T, Tardin C, Rousseau P, Salome L (2012) High-throughput single-molecule analysis of DNA-protein interactions by tethered particle motion. Nucleic Acids Research 40(12):e89–e89) for the detailed calculations of $R_{exp//raw}$ of the bead.

**Procedure of analysis for HT-TPM experimental data**

In order to quantify the small differences expected on $R_{exp//raw}$, we set up a two-step procedure that is described in detail below. Briefly, it consists in selecting traces fulfilling several criteria of validity and applying corrections for detector temporal averaging to their $R_{exp//raw}$. All this procedure was performed with homebuilt Mathematica scripts (available upon request).

*Criteria of validity of the DNA-bead complexes*

First, we discard the trajectories that have mean asymmetry factors above 1.35, calculated as the average of the asymmetric factors measured along the time trace, or that have a mean amplitude of motion smaller than 1 nm or higher than 1000 nm. Then the probability distribution of the average of for each trajectory is built with the remaining trajectories and fitted by a Gaussian distribution centered on a mean value, called mean with a standard deviation (sd). As we noticed that a few trajectories had $R_{exp//raw}$ averages standing out of the Gaussian distribution, we added a second step of validation to eliminate the misformed tethers with an average $R_{exp//raw}$ outside the interval (mean ± 2.5 sd). Using this criterion, no more than 1.3% of valid trajectories were eliminated during this additional step.

In total, about 12% trajectories were eliminated and the final number of valid trajectories eventually ranged between 348 and 3496 (See Table 1), depending on the DNA construct.

*Correction of time averaging effect*

Finite exposure time of detectors, $T_{ex}$, equal here to 40 ms, can lead to a blurring effect in single molecule (or particle) tracking experiments, as investigated for example in (Manghi M, et al. (2010) Probing DNA conformational changes with high temporal resolution by tethered particle motion. *Physical Biology* 7(4):046003). The correlation time of the positions, of about 20 ms, was calculated for each DNA-particle complex and then injected in Eq. 1 to correct the amplitudes of motion recorded on each trace.

$$R_{exp\|} = R_{exp\|raw} \left[ 2\frac{\tau_\|}{T_{ex}} - 2\left(\frac{\tau_\|}{T_{ex}}\right)^2 \left(1 - e^{-\frac{T_{ex}}{\tau_\|}}\right) \right]^{-1/2} \quad (1)$$

As previously, the trajectories with an average $R_{exp\|}$ outside the interval (mean ± 2.5 sd) were eliminated.

*Calculation of the amplitude of the motion of an ensemble of particles*

Lastly, the experimental value of the amplitude of the motion of an ensemble of particles was obtained by fitting the probability distribution of $R_{exp\parallel}$ with a Gaussian, which gave us its mean value as its center. The error on the amplitude of the motion of an ensemble of particles was calculated by using the bootstrap method of R software (R Foundation for Statistical Computing, Vienna, Austria). Doing so, we find a typical error of 0.4 nm (See Table 1).

**TPM simulations**

*DNA coarse-grained model*

We performed Kinetic Monte Carlo simulations (Newman MEJ, Barkema GT (1999) Monte Carlo Methods in Statistical Physics (Oxford University Press)) on the particle-DNA complex to predict the particle to anchor 2D-distance. The labeled DNA polymer is modeled as a chain of *N* connected small spheres of radius $a$, with a DNA contour length equal to $L = 2a(N-1)$, and a larger particle of radius $R_p$ = 150 nm $\geq a$ at its terminus. At this level of modeling, the internal structure of the double-stranded DNA is not considered and the persistence length, $L_p$, is averaged over the nucleotide sequence and taken equal to 147 bp. To model the angle imposed in the experiments by the *n* successive 10-bp CA$_6$CGG inserts, we have incorporated a bend of fixed angle $\theta$ between the three small beads located in the middle of the DNA molecule, which is simulated using a coarse-grained model. More precisely, the elastic term between triplets *(i−1, i, i+1)* of successive beads, making an angle $\theta_i$, namely $U_i = \kappa(1 - cos\,\theta_i)$, is replaced by $U'_i = \kappa(1 - cos(\theta_i - \theta))$ for the central triplet where $\kappa$ is the bending modulus and $\kappa = 147 k_B T$. The polymer is grafted on a surface. Since the polymer motion is limited to the upper half-plane, we impose a "hard wall" boundary condition for monomer spheres and for the particle. All spheres interact via stretching and bending forces. We invite the reader to refer to (Manghi M, et al. (2010) Probing DNA conformational changes with high temporal resolution by tethered particle motion. Physical Biology 7(4):046003) for the details of these Kinetic Monte Carlo simulations.

*Simulated bent DNA*

The bent sequences used in the experiments are simulated by setting a fixed angle between three successive monomers located at the center of the DNA molecule. A full range of angles were studied in successive simulations: 0, 18, 30, 45, 50, 60, 72, 90, 120 and 180 degrees.

*Extraction/Calculation of the particle to anchor 2D-distance of the simulated particle*

The 2D-vector of the particle position $\vec{R}_{sim\parallel}$ is measured throughout simulations and utilized to estimate the amplitude of motion defined as $\sqrt{\langle \vec{R}^2_{sim\parallel} \rangle}$, the average being taken along the trajectory.

## Theory for a local bend

We consider a homogeneous polymer of length $L$ using the Worm-Like Chain (WLC) model by Kratky and Porod with a persistence length $L_p$ (Kratky O, Porod G (1949) Röntgenuntersuchung gelöster Fadenmoleküle. Recl Trav Chim Pays-Bas 68(12):1106–1122). We suppose that a kink is located at distance $l$ from one end and locally induces a spontaneous curvature with an angle $\theta$, see below.

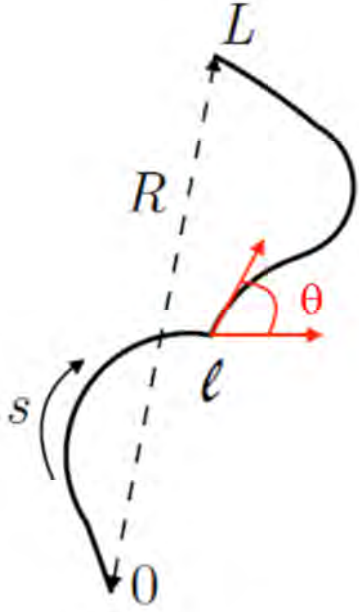

We denote the tangent vector at curvilinear position $s$ by $\vec{t(s)}$, and the tangent-tangent correlation function is given by $\langle \vec{t(s)} \cdot \vec{t(s')} \rangle = exp[-|s - s'|/L_P]$.

The mean-square end-to-end distance is defined as: $\langle R^2 \rangle = \int_0^L ds \int_0^L ds' \, \langle \vec{t(s)} \cdot \vec{t(s')} \rangle$.

Integration leads to the following result:

$$\langle R^2 \rangle = 2\, L_P^2 \left[ \frac{L}{L_P} - 2 + e^{-\frac{l}{L_P}} + e^{-\frac{L-l}{L_P}} + cos(\theta) \left( 1 - e^{-\frac{L-l}{L_P}} - e^{-\frac{l}{L_P}} + e^{-\frac{L}{L_P}} \right) \right]$$

## Theory for a local stiffer insert

We use the same model and assume now that a stiffer insert of length $b$ and persistence length $L_{Pb}$ is inserted at position $l$, see below:

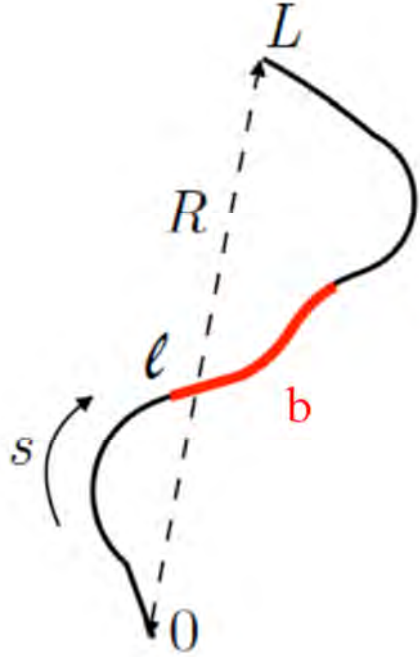

The mean-square end-to-end distance is again $\langle R^2 \rangle = \int_0^L ds \int_0^L ds' \langle \vec{t(s)} \cdot \vec{t(s')} \rangle$. To take into account the 3 parts of the DNA molecule, the calculation requires the equation to be cut into parts. After a straightforward integration, one gets:

$$\langle R^2 \rangle = 2\, L_P^2 \left[ \frac{L-b}{L_P} + e^{-\frac{l}{L_P}} + e^{-\frac{L-b-l}{L_P}} - 2 + e^{-\frac{b}{L_{Pb}}} \left(1 - e^{-\frac{l}{L_P}}\right)\left(1 - e^{-\frac{L-b-l}{L_P}}\right) \right.$$

$$\left. + 2 L_P L_{Pb} \left(1 - e^{-\frac{b}{L_{Pb}}}\right)\left(2 - e^{-\frac{l}{L_P}} - e^{-\frac{L-b-l}{L_P}}\right) \right] + 2 L_{Pb}^2 \left(\frac{b}{L_{Pb}} - 1 + e^{-\frac{b}{L_{Pb}}}\right)$$

**S1 figure:** linearity of $R_{DNA}$ with the length of a rigid insert. Simulated data are represented by black markers and the linear fits by red straight lines.

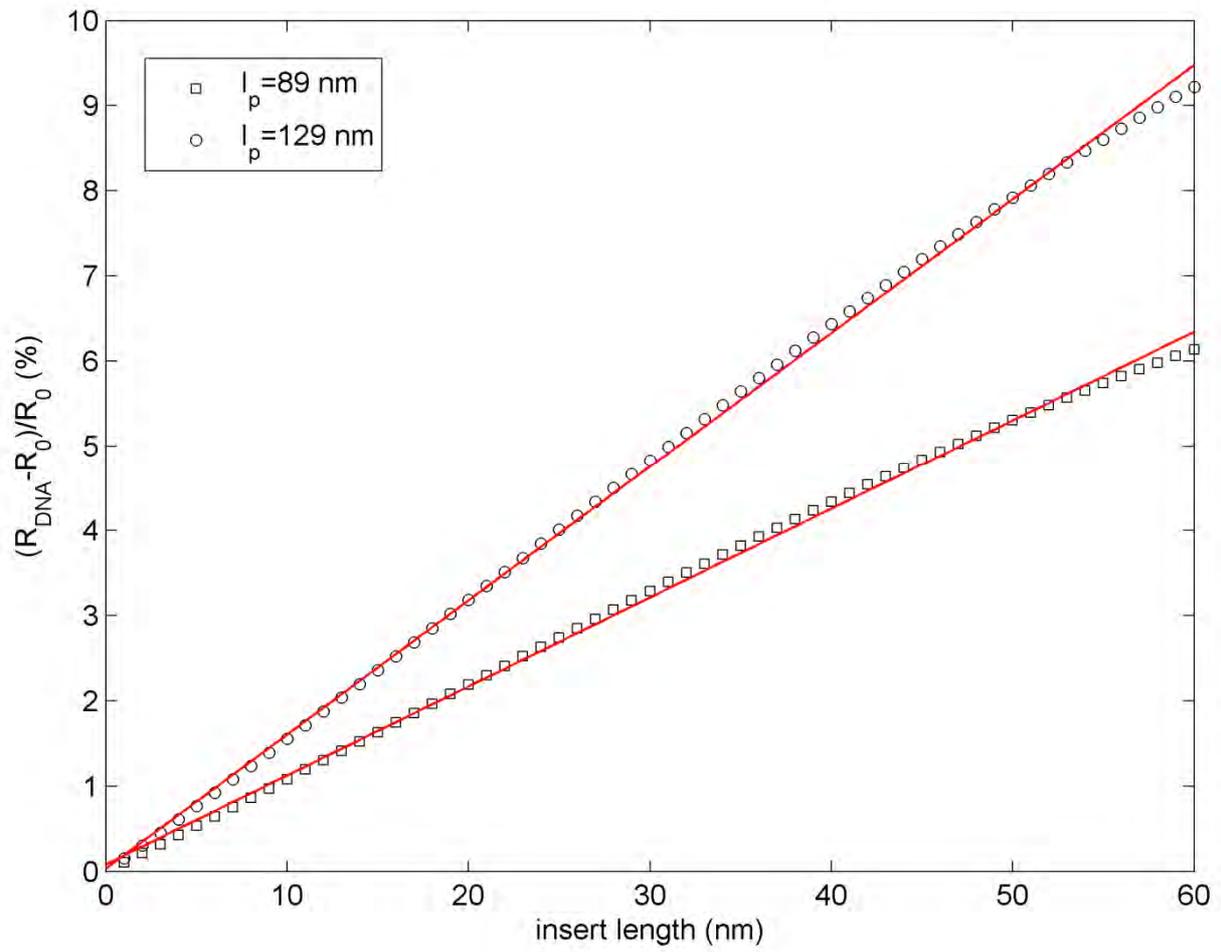

**S2 figure:** complement to Fig. 3

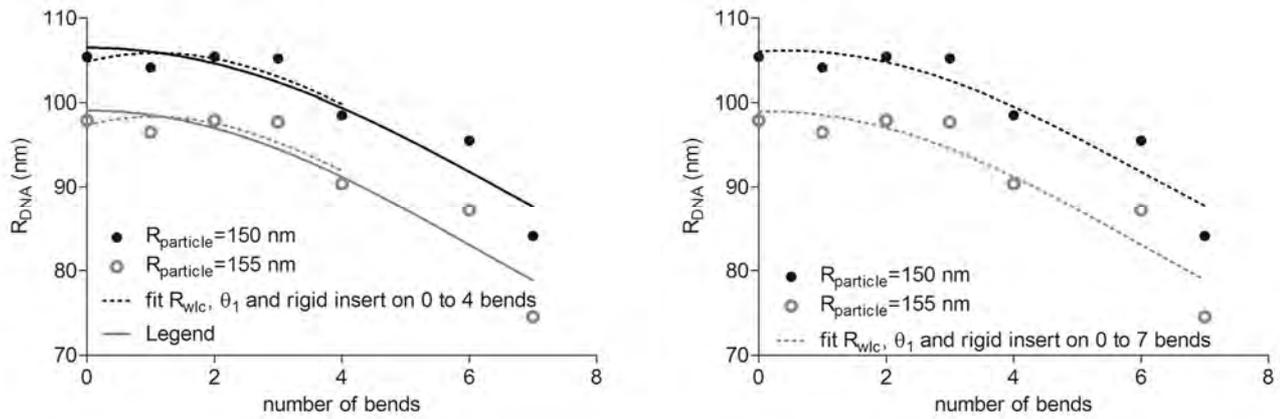

Experimental data (symbols) obtained after deconvolution of the particle effects using the minimal model and considering the size of labelling particle equal to 150 nm (●) and 155 nm (○), and two series of fits with $R_{WLC}$ and $\theta_1$ (–) (already shown in figure 3), or $R_{WLC}$, $\theta_1$ and a rigidity term $\alpha$ (…) as free parameters. The results are shown in S3 Table.

**S3 table:** complement to S2

$$\sqrt{\langle R^2_{WLC}\rangle} = D[(1 + 0.338\, cos(n\theta_1 - \theta_0)\,]^{1/2} + \alpha n$$

| Fit conditions | $R_p$ | D (nm) | $\theta_1$ (°) | $\theta_0$ (°) | $\alpha$ (nm) |
|---|---|---|---|---|---|
| No global curvature ($\theta_0 = 0$) No local rigidity ($\alpha = 0$) | 150 | 92 ± 2 | 15 ± 2 | | |
| | 155 | 86 ± 2 | 17 ± 2 | | |
| No local rigidity ($\alpha = 0$) | 150 | 91 ± 2 | 19 ± 4 | 25 ± 19 | |
| | 155 | 95 ± 2 | 21 ± 4 | 29 ± 22 | |
| No global curvature ($\theta_0 = 0$) Data fitted from n =0 to 4 | 150 | 91 ± 2 | 17 ± 12 | | 2 ± 4 |
| | 155 | 84 ± 3 | 22 ± 15 | | 2 ± 4 |
| No global curvature ($\theta_0 = 0$) Data fitted from n =0 to 7 | 150 | 92 ± 3 | 17 ± 12 | | 1 ± 4 |
| | 155 | 86 ± 3 | 17 ± 15 | | 0 ± 4 |

**S4 figure: non-linearity of $R_{DNA}$ with the persistence length of a rigid insert.**

$R_{DNA}$ is computed considering: $L=575$ bp, $l=274$ bp, $b=88$ bp and with $L_p=150$ bp outside the 88bp insert. As a result, $R_{DNA}$ appears to be poorly sensitive to an increase of rigidity of the insert for persistence lengths of the insert > 150 bp.

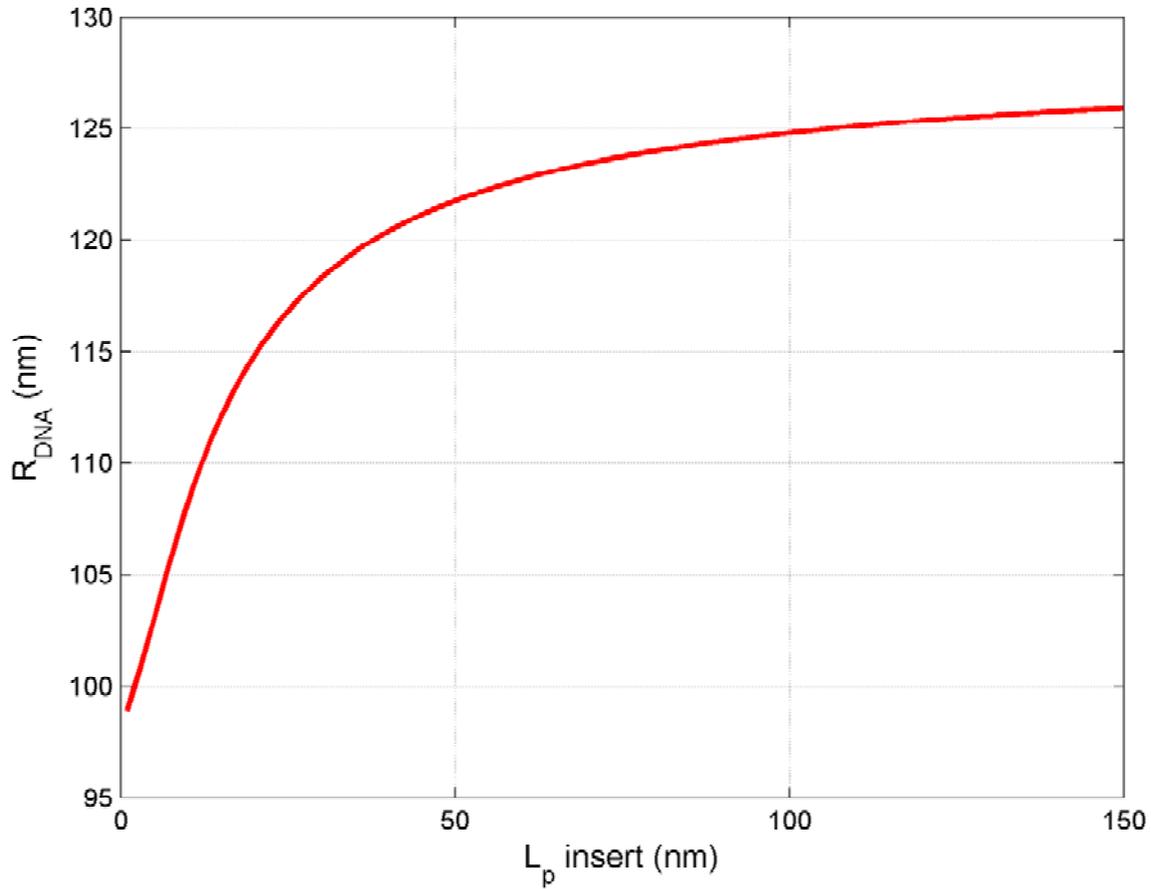

**S5 table**

| DNA test | DNA reference | $R_{DNAtest}$ | $R_{DNAref}$ | $\dfrac{[\langle R^2_{DNA}\rangle_{REF} - \langle R^2_{DNA}\rangle_{TEST}]}{\langle R^2_{DNA}\rangle_{REF}}$ | θ (°) | $\dfrac{\theta\ (°)}{n}$ |
|---|---|---|---|---|---|---|
| 6An4$^P$ | 6An0 | 98.5 | 105.4 | 0.127 | 60 | 15 |
| 6An6$^P$ | 6An0 | 95.5 | 105.4 | 0.179 | 73 | 12 |
| 6An7$^P$ | 6An0 | 84.1 | 105.4 | 0.363 | 116 | 17 |
| 6An4$^O$ | 6An0 | 108.6 | 105.4 | -0.062 | - | - |

# Video S1

Typical field of Particle-DNA complexes recorded by HT-TPM (area size about 100 μm X150 μm)

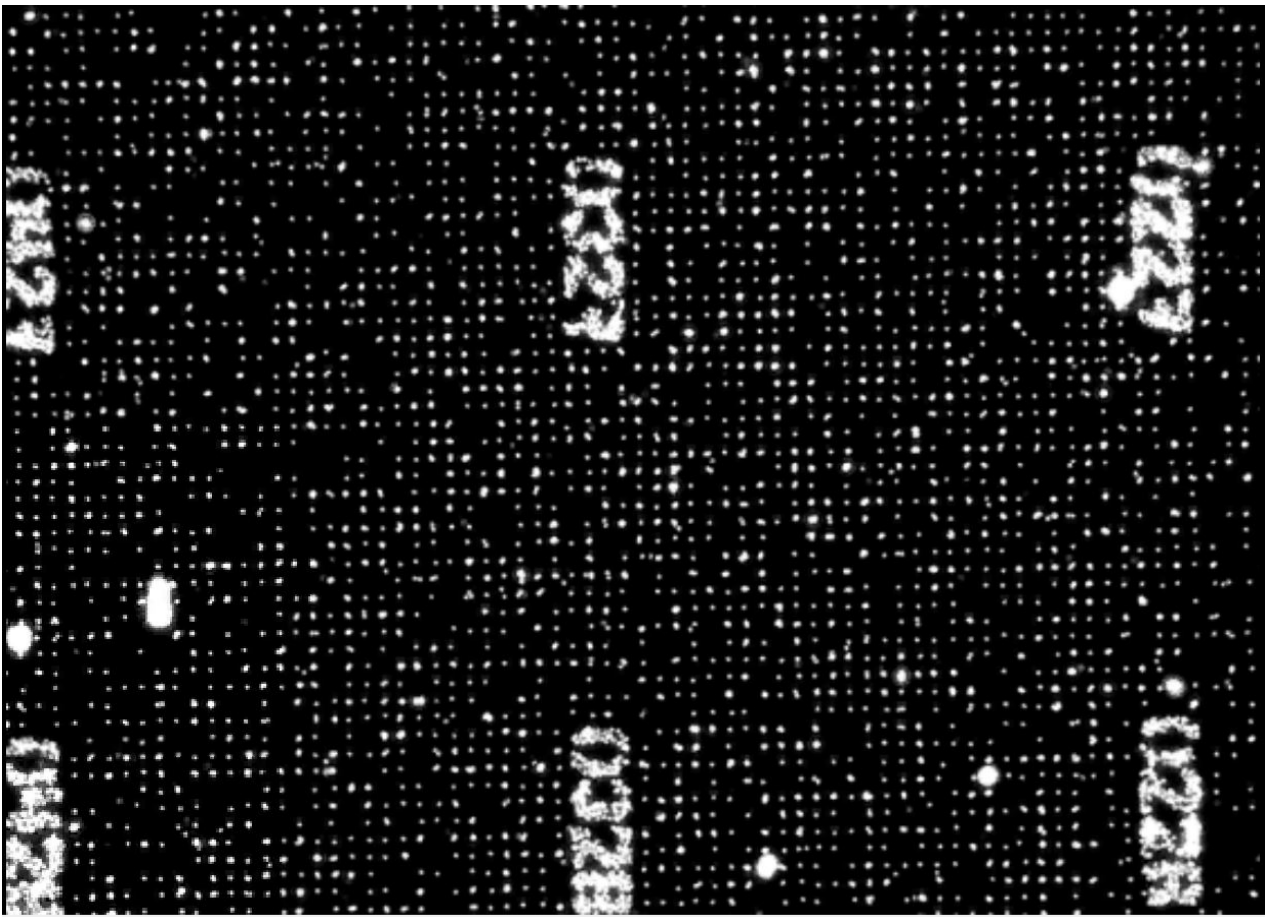